\documentclass[aps,pra,twocolumn,showpacs,preprintnumbers,amsmath,amssymb]{revtex4}
\usepackage{dcolumn}
\usepackage{bm}

\begin{document}


\title{On expansion of equal-time relativistic two-body wave equations in powers of $1/c$ \\ to higher orders}

\author{Alexei Turovsky}
   \email{turovsky@bitp.kiev.ua}
\affiliation{Bogolyubov Institute for Theoretical Physics, Kyiv 03680, Ukraine}
\date{\today}

\begin{abstract}
Based on an extension of the Foldy--Wouthuysen method to two-body equations, the problem of expansion of equal-time relativistic equations for two Dirac particles in powers of $1/c$ to higher orders is considered. For the case of two particles with unequal masses, the transformed Hamiltonian in a general even-even form is obtained to order $1/c^4$. It is found that certain extra terms, which can be eliminated by an additional unitary transformation, arise in the expansion in higher orders, depending on the order of application of the generating functions in the first iteration. As examples for illustration, the Breit equation and the Salpeter equation with the Breit interaction are taken and their reduction to approximate forms including all the $1/c^{4}$-order terms is carried out using the method under consideration. The obtained results may be applied for the nonrelativistic expansion of two-body wave equations with various interaction potentials to higher orders, for the investigation of their features and symmetries, and may also be useful in the study of light atoms.
\end{abstract}


\maketitle

\section{\label{sec:1}Introduction}
In spite of considerable achievements in quantum field theory, particularly in quantum electrodynamics (QED), relativistic and semirelativistic two-body wave equations are widely used to learn about various relativistic effects in quantum systems and also to study and calculate their energy spectra (see, e.g., \cite{list}). In many cases, however, to obtain an acceptable description of a two-body system, it is not essential to solve the original equations of motion and one can restrict oneself to the consideration of their approximate forms with a small parameter, such as the inverse speed of light. As a rule, for many applications it is sufficient to restrict oneself to an expansion of the wave equations in powers of $1/c$ to second order. Thus, e.g., the well-known Breit-Fermi Hamiltonian and its special case for a vector exchange, which can be derived from the nonrelativistic reduction of different two-body equations, are often used to describe spectra of atomic, nuclear, and quark systems \cite{BetheSalpeter,green,lucha}.

A relatively simple description of a system of two spin-half particles, including relativistic effects and its energy spectrum, can be given with the help of equations of the Breit type. In order to expand the equations, the Foldy--Wouthuysen (FW) method \cite{bjorkendrell} generalized by Chraplyvy to the two-body problem is usually applied \cite{chraplyvy,chraplyvy2,barker,chraplyvy3,eriksen,eriksen2}.

According to this method, one represents a relativistic two-body Hamiltonian in such a general form:
\begin{equation}\label{Eq:1}
   H=\beta_1m_1c^2+\beta_2m_2c^2+(\mathcal{EE})+(\mathcal{EO})+(\mathcal{OE})+(\mathcal{OO}).
\end{equation}
This is a sum of the two ``large'' terms $\beta_1m_1c^2+\beta_2m_2c^2$ and even-even, even-odd, odd-even, and odd-odd terms, respectively. They are, in fact, matrices of $16\times16=256$ elements, and can be written as direct products of four-by-four matrices of each particle. In order to remove the undesirable terms (even-odd, etc.), the Hamiltonian must be subjected to canonical transformations of the type
\begin{eqnarray}\label{Eq:2}
   e^{iS}He^{-iS}&=&H+i[S,H]+\frac{(i)^2}{2!}[S,[S,H]] \nonumber \\ \nonumber \\
      &&+\;\frac{(i)^3}{3!}[S,[S,[S,H]]]+\cdots,
\end{eqnarray}
provided the generating functions $S$ are suitably-chosen Hermitian operators, and where the inverse speed of light (or, equivalently, the inverse masses) can be taken as an expansion parameter.

Though in the case of interacting particles one cannot determine a unitary transformation that cancels all the undesirable terms in all orders in principle, the original Hamiltonian can be converted by the procedure (\ref{Eq:2}) into an even-even operator to any desired accuracy. There are a lot of sets of $S$ \cite{chraplyvy,chraplyvy2} to reduce a two-body Hamiltonian to approximate forms. In this paper, we use the generators in a simple form, which, for the first iteration, read as follows:
\begin{subequations}\label{Eq:3}
\begin{eqnarray}
   S_{oe}&=&-\frac{i\beta_1}{2m_1c^2}(\mathcal{OE}),   \label{Eq:3a}\\ \nonumber\\
   S_{eo}&=&-\frac{i\beta_2}{2m_2c^2}(\mathcal{EO}),   \label{Eq:3b}\\ \nonumber\\
   S_{oo}&=&-\frac{i(\beta_1m_1-\beta_2m_2)}{2(m_1^2-m_2^2)c^2}(\mathcal{OO}).  \label{Eq:3c}
\end{eqnarray}
\end{subequations}
The generating functions for the next iterations differ from those in Eq. (\ref{Eq:3}) by the structure of their odd-even, even-odd, and odd-odd factors. These generators enable one to reduce a sixteen-component two-body equation to its four-component approximate forms relative to the chosen energy states of the two particles. They were first introduced by Chraplyvy in \cite{chraplyvy} to convert the two-body Hamiltonian (\ref{Eq:1}) into an even-even operator to order $1/c^{2}$ (under the assumption that $(\mathcal{EE})$, $(\mathcal{OO})$ are of order $c^0$, and $(\mathcal{OE})$, $(\mathcal{EO})$ of order $c^1$),
\begin{subequations}\label{Eq:4}
\begin{eqnarray}
  H_{\rm tr}&\approx&\beta_1m_1c^2+\beta_2m_2c^2+(\mathcal{EE})    \label{Eq:4a}
  \\ \nonumber \\
     &&+\frac{\beta_1}{2m_1c^2}(\mathcal{OE})^2+\frac{\beta_2}{2m_2c^2}(\mathcal{EO})^2    \label{Eq:4b}
  \\ \nonumber \\
     &&+\frac{1}{8m_1^2c^4}[(\mathcal{OE}),[(\mathcal{EE}),(\mathcal{OE})]]   \nonumber
  \\ \nonumber\\
     &&~~~~~~~~+\frac{1}{8m_2^2c^4}[(\mathcal{EO}),[(\mathcal{EE}),(\mathcal{EO})]]   \label{Eq:4c}
  \\ \nonumber \\
     &&-\frac{\beta_1}{8m_1^3c^6}(\mathcal{OE})^4-\frac{\beta_2}{8m_2^3c^6}(\mathcal{EO})^4   \label{Eq:4d}
  \\ \nonumber\\
     &&+\frac{\beta_1\beta_2}{8m_1m_2c^4}\Bigl\{[(\mathcal{OE}),[(\mathcal{EO}),(\mathcal{OO})]_{+}]_{+}  \nonumber
  \\ \nonumber\\
            &&~~~~~~~~
            +[(\mathcal{EO}),[(\mathcal{OE}),(\mathcal{OO})]_{+}]_{+}
         \Bigr\}         \label{Eq:4e}
  \\ \nonumber\\
     &&+\frac{\beta_1m_1-\beta_2m_2}{2(m_1^2-m_2^2)c^2}(\mathcal{OO})^2   \label{Eq:4f}
  \\ \nonumber\\
     &&+\frac{\beta_2m_1-\beta_1m_2}{8m_1m_2(m_1^2-m_2^2)c^6}[(\mathcal{OE}),(\mathcal{EO})]^2   \label{Eq:4g}
  \\ \nonumber\\
     &&-\frac{\beta_1m_1+\beta_2m_2}{16m_1^2m_2^2c^6}[(\mathcal{OE})^2,(\mathcal{EO})^2]_+       \label{Eq:4h}
  \\ \nonumber\\
     &&+\frac{\beta_1}{8m_1m_2^2c^6}(\mathcal{EO})(\mathcal{OE})^2(\mathcal{EO})   \nonumber
  \\ \nonumber\\
         &&~~~~~~~~+\frac{\beta_2}{8m_1^2m_2c^6}(\mathcal{OE})(\mathcal{EO})^2(\mathcal{OE})     \label{Eq:4i}
  \\ \nonumber\\
     &&+\frac{\beta_1\beta_2(m_1^2+m_2^2)-2m_1m_2}{8m_1m_2(m_1^2-m_2^2)c^4}                      \nonumber
  \\ \nonumber\\
         &&~~~~~~~~\times[[(\mathcal{EO}),(\mathcal{OE})],(\mathcal{OO})]\,.          \label{Eq:4j}
\end{eqnarray}\end{subequations}
Because of the form of (\ref{Eq:3c}) this prescription is applicable only for the case of unequal masses, however, it enables one to get four forms of the reduced Hamiltonian relative to the four different energy states of the system.

Equation (\ref{Eq:4}) was applied by Barker and Glover to the Coulomb case in order to transform the Hermitian part of the three-dimensional Bethe-Salpeter equation (the Salpeter equation \cite{salpeter}) and the Breit equation to sixteen-component approximate forms to order $1/c^2$ in \cite{barker}, where a comparative analysis of the transformed Hamiltonians for these equations was carried out and significant differences between them were noted.

Yet, a certain further expansion of the relativistic two-body wave equations to higher orders in $1/c$ may be of some interest as well, and not only from the theoretical or mathematical point of view. The improvement of experimental techniques for the study of energy spectra in atomic systems, first of all, in hydrogen-like atoms, muonium, and positronium, allows one to make precise measurements of their energy levels \cite{sokolov,mohr,karshenboim}. They usually apply QED to learn about the $\alpha^5mc^2$, $\alpha^6mc^2$ corrections to the energy (see, e.g., \cite{douglaskroll,pachucki1996,darewych,pachucki,pachuckiyerokhin,mart,elekina,jentschura,skarshenboim} and references therein), however, the problem of derivation of the higher-order Hamiltonian of an arbitrary light atom remains difficult. Nevertheless, an expansion of relativistic two-body wave equations to order $1/c^4$ provides a rather straightforward derivation of an effective $\alpha^6mc^2$ Hamiltonian, though it gives incomplete treatment of relativistic and quantum field effects. Still, it enables us to get some information about the structure of terms contributing to $\alpha^6mc^2$ for the energy eigenvalues. The nonrelativistic expansion to higher orders can also help to investigate the equations of motion themselves, their features and symmetries, to answer questions on relativistic effects they involve, what energies they work to, and so on.

This paper focuses on the problem of expansion of the equal-time relativistic wave equations for two Dirac particles to order $1/c^4$ by means of the extension of the FW method to two-body systems, using the generating functions (\ref{Eq:3}), and is organized as follows. Section~\ref{sec:2} deals with the two-body Hamiltonian transformed to higher orders, for which all the $1/c^4$-order terms are found, and which is a continuation of the expansion (\ref{Eq:4}) in the case of commutation of $(\mathcal{OE})$ and $(\mathcal{EO})$. It occurs that the form of the higher-order part of the transformed Hamiltonian depends on the order of application of the functions (\ref{Eq:3}), namely, it can involve certain extra terms having mass differences in the denominators even if both particles are in a positive energy state. Here, we are concerned with an additional unitary transformation canceling terms of this type as well.

In Section~\ref{sec:3}, we work out all the terms of order $1/c^4$ in the expansion of the Breit equation and the Salpeter equation with the Breit interaction. In addition, we also discuss the possibility of modification of the $1/c^4$-order part of the transformed Hamiltonians with the help of unitary transformations. Finally, Section~\ref{sec:4} contains a summary and main conclusions of this article.

\section{\label{sec:2}Hamiltonian transformed to higher orders}
Proceeding with the procedure of transformation of the Hamiltonian (\ref{Eq:1}), using the generators (\ref{Eq:3}) and with due regard for the commutation relation
\begin{equation}\label{Eq:5}
   [(\mathcal{OE}),(\mathcal{EO})]=0,
\end{equation}
which causes a considerable simplification in $H_{\rm tr}$, we get new even-even terms coming after the terms written in Eq.~(\ref{Eq:4}), and which are of lower order of magnitude. They form the higher-order, with respect to $1/c$, part of the transformed Hamiltonian.
\begin{widetext}
Thus one obtains such a prescription for the transformation of $H$:
\begin{subequations}\label{Eq:6}
\begin{eqnarray}
   H_{\rm tr}&\approx&\beta_1m_1c^2+\beta_2m_2c^2+(\mathcal{EE})
        +\frac{\beta_1}{2m_1c^2}(\mathcal{OE})^2
        +\frac{\beta_2}{2m_2c^2}(\mathcal{EO})^2
        +\frac{\beta_1m_1-\beta_2m_2}{2(m_1^2-m_2^2)c^2}(\mathcal{OO})^2
   \nonumber \\ \nonumber \\
      &&+\frac{1}{8m_1^2c^4}[(\mathcal{OE}),[(\mathcal{EE}),(\mathcal{OE})]]
        +\frac{1}{8m_2^2c^4}[(\mathcal{EO}),[(\mathcal{EE}),(\mathcal{EO})]]
        +\frac{\beta_1\beta_2}{4m_1m_2c^4}[(\mathcal{OE}),[(\mathcal{EO}),(\mathcal{OO})]_{+}]_{+}
   \nonumber \\ \nonumber\\
      &&-\frac{\beta_1}{8m_1^3c^6}(\mathcal{OE})^4-\frac{\beta_2}{8m_2^3c^6}(\mathcal{EO})^4
   \label{Eq:6a} \\ \nonumber \\
      &&-\frac{\beta_1}{8m_1^3c^6}[(\mathcal{OE}),(\mathcal{EE})]^2
        -\frac{\beta_2}{8m_2^3c^6}[(\mathcal{EO}),(\mathcal{EE})]^2
   \label{Eq:6b} \\ \nonumber\\
      &&+\frac{\beta_1}{8m_1^{}m_2^2c^6}[(\mathcal{EO}),(\mathcal{OO})]_{+}^2
        +\frac{\beta_2}{8m_1^2m_2^{}c^6}[(\mathcal{OE}),(\mathcal{OO})]_{+}^2
   \label{Eq:6c} \\ \nonumber\\
      &&-\frac{\beta_1m_1-\beta_2m_2}{16m_1^2(m_1^2-m_2^2)c^6}
           [(\mathcal{OO}),[(\mathcal{OE}),[(\mathcal{OE}),(\mathcal{OO})]_+]_+]_+
        -\frac{\beta_1(\beta_1m_1-\beta_2m_2)^2}{16m_1(m_1^2-m_2^2)^2c^6}
           [(\mathcal{OO}),[(\mathcal{OO}),(\mathcal{OE})^2]_+]_+
   \nonumber \\ \nonumber\\
      &&-\frac{\beta_1m_1-\beta_2m_2}{16m_2^2(m_1^2-m_2^2)c^6}
           [(\mathcal{OO}),[(\mathcal{EO}),[(\mathcal{EO}),(\mathcal{OO})]_+]_+]_+
        -\frac{\beta_2(\beta_1m_1-\beta_2m_2)^2}{16m_2(m_1^2-m_2^2)^2c^6}
           [(\mathcal{OO}),[(\mathcal{OO}),(\mathcal{EO})^2]_+]_+ \qquad
   \label{Eq:6d} \\ \nonumber\\
      &&-\frac{\beta_1m_2-\beta_2m_1}{8m_1m_2(m_1^2-m_2^2)c^6}
           [(\mathcal{OO}),[(\mathcal{OE}),[(\mathcal{EO}),(\mathcal{EE})]]]_+
   \nonumber \\ \nonumber\\
      &&+\frac{\beta_1}{8m_1^{}m_2^2c^6}
           [[(\mathcal{EO}),(\mathcal{EE})],[(\mathcal{OE}),(\mathcal{OO})]_+]
        +\frac{\beta_2}{8m_1^2m^{}_2c^6}
           [[(\mathcal{OE}),(\mathcal{EE})],[(\mathcal{EO}),(\mathcal{OO})]_+]
   \label{Eq:6e} \\ \nonumber\\
      &&+\frac{(\beta_1m_1-\beta_2m_2)^2}{8(m_1^2-m_2^2)^2c^4}
           [(\mathcal{OO}),[(\mathcal{EE}),(\mathcal{OO})]]
   \label{Eq:6f}\\ \nonumber\\
      &&+\frac{1}{384m_1^4c^8}\Bigl\{
           [(\mathcal{OE}),[(\mathcal{OE}),[(\mathcal{OE}),[(\mathcal{OE}),(\mathcal{EE})]]]]
           +32[(\mathcal{OE})^3,[(\mathcal{OE}),(\mathcal{EE})]]
        \Bigr\}
   \nonumber \\ \nonumber\\
      &&+\frac{1}{384m_2^4c^8}\Bigl\{
           [(\mathcal{EO}),[(\mathcal{EO}),[(\mathcal{EO}),[(\mathcal{EO}),(\mathcal{EE})]]]]
           +32[(\mathcal{EO})^3,[(\mathcal{EO}),(\mathcal{EE})]]
         \Bigr\}
   \label{Eq:6g}\\ \nonumber\\
      &&+\frac{1}{64m_1^2m_2^2c^8}
           [(\mathcal{OE}),[(\mathcal{OE}),[(\mathcal{EO}),[(\mathcal{EO}),(\mathcal{EE})]]]]
   \label{Eq:6h}\\ \nonumber\\
      &&-\frac{\beta_1\beta_2}{96m_1^3m_2^{}c^8}\Bigl\{
           [(\mathcal{OE}),[(\mathcal{OE}),[(\mathcal{OE}),[(\mathcal{EO}),
              (\mathcal{OO})]_+]_+]_+]_+
           +8[(\mathcal{OE})^3,[(\mathcal{EO}),(\mathcal{OO})]_+]_+
         \Bigr\}
   \nonumber \\ \nonumber\\
      &&-\frac{\beta_1\beta_2}{96m_1^{}m_2^3c^8}\Bigl\{
           [(\mathcal{OE}),[(\mathcal{EO}),[(\mathcal{EO}),[(\mathcal{EO}),
               (\mathcal{OO})]_+]_+]_+]_+
           +8[(\mathcal{OE}),[(\mathcal{EO})^3,(\mathcal{OO})]_+]_+
         \Bigr\}
   \label{Eq:6i} \\ \nonumber\\
      &&+\frac{\beta_1}{16m_1^5c^{10}}(\mathcal{OE})^6
        +\frac{\beta_2}{16m_2^5c^{10}}(\mathcal{EO})^6.
   \label{Eq:6j}
\end{eqnarray}
\end{subequations}
\end{widetext}
The FW method allows one to expand the Hamiltonian to any desired degree of approximation, keeping its Hermitian character. Here, we assume that $(\mathcal{EE})$, $(\mathcal{OO})$ are of order $c^0$, and $(\mathcal{OE})$, $(\mathcal{EO})$ of order $c^1$, and we retain the terms up to order $1/c^4$. So, under this assumption, expression (\ref{Eq:6}) represents the transformed Hamiltonian approximate out to fourth order. It can be divided into two parts. The first part consists of the terms (\ref{Eq:6a}) and is the transformed Hamiltonian to order $1/c^2$, into which the expression (\ref{Eq:4}) goes over under the commutation relation (\ref{Eq:5}). The second part, which we will refer to as the higher-order transformed Hamiltonian, involves all the $1/c^4$-order terms (\ref{Eq:6b}...j). It is a sixteen-component equation, and as usual, to obtain its four-component forms, i.e. reduced Hamiltonians, one has to put $\beta_1=\beta_2=\pm1$ or $\beta_1=-\beta_2=\pm1$. Note that if the inverse mass is considered as an expansion parameter, the terms (\ref{Eq:6b}...e), which are nonlinear in $(\mathcal{E}\mathcal{E})$, $(\mathcal{OO})$, are of order $1/m^3$. Many of them have a mass difference in the denominators. The terms (\ref{Eq:6g}, h, i), which are linear in $(\mathcal{EE})$, $(\mathcal{OO})$, are of order $1/m^4$. The term (\ref{Eq:6f}), which consists of the even-even and odd-odd operators, is the only one of order $1/m^2$ in the higher-order transformed Hamiltonian.

For the case when the mass of one particle becomes considerably great in compare with the mass of the other one, i.e. $m_1\rightarrow\infty$ or $m_2\rightarrow\infty$, equation (\ref{Eq:6}) goes over into the corresponding formula for a single Dirac particle in external fields:
\begin{eqnarray}\label{Eq:7}
   H_{\rm tr}&=&\beta mc^2+\mathcal{E}+\frac{\beta}{2mc^2}\mathcal{O}^2
              +\frac{1}{8m^2c^4}[\mathcal{O},[\mathcal{E},\mathcal{O}]]
   \nonumber \\ \nonumber \\
            &&-\frac{\beta}{8m^3c^6}\mathcal{O}^4
              -\frac{\beta}{8m^3c^6}[\mathcal{O},\mathcal{E}]^2
   \nonumber \\ \nonumber \\
            &&+\frac{1}{384m^4c^8}
                 [\mathcal{O},[\mathcal{O},[\mathcal{O},[\mathcal{O},\mathcal{E}]]]]
   \nonumber \\ \nonumber \\
            &&+\frac{1}{12m^4c^8}[\mathcal{O}^3,[\mathcal{O},\mathcal{E}]]
              +\frac{\beta}{16m^5c^{10}}\mathcal{O}^6,\quad
\end{eqnarray}
which is the transformation of the Hamiltonian
\begin{equation*}
   H=\beta mc^2+\mathcal{E}+\mathcal{O},
\end{equation*}
with the use of the generator
\begin{equation*}
   S=-\frac{i\beta}{2mc^2}\,\mathcal{O}.
\end{equation*}
Under this condition only the terms of (\ref{Eq:6b},~g,~j) remain in the higher-order transformed Hamiltonian and pass into the last four terms in Eq.~(\ref{Eq:7}); the rest of the commutators and anticommutators which involve the $(\mathcal{OO})$ terms in the expression (\ref{Eq:6}) vanish.

The transformed Hamiltonian written out in Eq.~(\ref{Eq:4}) to second order is derived with the use of the generating functions (\ref{Eq:3}) and it is not very important which of them is used first in the first iteration in the series (\ref{Eq:2}); regardless of their order of application in this iteration, the same expression in the form of (\ref{Eq:4}) will be obtained. In other words, it is not important which of the undesirable terms in Eq. (\ref{Eq:1}) will be destroyed first. But this statement is correct only if one needs to get the expansion up to the terms written in Eq.~(\ref{Eq:4}), or the same, to second order in $1/c$ under our assumption. The use of $S_{oe}$ or $S_{eo}$ first in the sequence in the first iteration leads to the transformed Hamiltonian in the form (\ref{Eq:6}), however, if one applies $S_{oo}$ first (instead of $S_{oe}$ or $S_{eo}$) in (\ref{Eq:2}), certain ``extra'' terms of fourth order will arise in the transformed Hamiltonian in addition to those in Eq.~(\ref{Eq:6}); namely,
\begin{eqnarray*}
   &&\frac{\beta_1m_2-\beta_2m_1}{8m_1m_2(m_1^2-m_2^2)c^6}
       [[(\mathcal{OE})(\mathcal{EO}),(\mathcal{OO})],(\mathcal{EE})]
   \\ \\
   &&+\frac{\beta_1m_2+\beta_2m_1}{8m_1m_2(m_1^2-m_2^2)c^6}
   \\ \\
   &&\quad\times[(\mathcal{EO})(\mathcal{OO})(\mathcal{OE})
                   -(\mathcal{OE})(\mathcal{OO})(\mathcal{EO}),(\mathcal{EE})]
\end{eqnarray*}
\begin{eqnarray}\label{Eq:8}
   &&+\frac{m_2-\beta_1\beta_2m_1}{16m_1^2m_2(m_1^2-m_2^2)c^8}
       [[(\mathcal{OE})(\mathcal{EO}),(\mathcal{OO})],(\mathcal{OE})^2]            \nonumber \\ \nonumber \\
   &&+\frac{\beta_1\beta_2m_2-m_1}{16m_1m_2^2(m_1^2-m_2^2)c^8}
       [[(\mathcal{OE})(\mathcal{EO}),(\mathcal{OO})],(\mathcal{EO})^2]            \nonumber \\ \nonumber \\
   &&+\frac{m_2+\beta_1\beta_2m_1}{16m_1^2m_2(m_1^2-m_2^2)c^8}                     \nonumber \\ \nonumber \\
   &&\quad\times[(\mathcal{EO})(\mathcal{OO})(\mathcal{OE})
                   -(\mathcal{OE})(\mathcal{OO})(\mathcal{EO}),(\mathcal{OE})^2]   \nonumber \\ \nonumber \\
   &&+\frac{\beta_1\beta_2m_2+m_1}{16m_1m_2^2(m_1^2-m_2^2)c^8}                     \nonumber \\ \nonumber \\
   &&\quad\times[(\mathcal{EO})(\mathcal{OO})(\mathcal{OE})
                   -(\mathcal{OE})(\mathcal{OO})(\mathcal{EO}),(\mathcal{EO})^2].
\end{eqnarray}
As follows from this equation, the terms with mass differences in the denominators appear even though both particles are in positive or negative energy states (those correspond to setting $\beta_1=\beta_2=\pm1$). Such terms, but with other numerical factors, also appear in $H_{\rm tr}$ if one takes the sum $S=S_{oe}+S_{eo}+S_{oo}$ as a generating function in the series (\ref{Eq:2}). Obviously, some extra terms enter into the expansion and in sixth order and higher. This dependence of the form for the higher-order part of the transformed Hamiltonian on the order of application of the generators may be an important feature of the expansion to higher orders, and makes the difference between the second-order and higher-order expansions.

Still, the extra terms (\ref{Eq:8}) can be eliminated given that the transformed Hamiltonian is subjected to an additional unitary transformation like (\ref{Eq:2}) with a generating function in the form of a Hermitian even-even operator, which we represent in such a manner:
\begin{subequations}\label{Eq:9}
\begin{eqnarray}
   S_{ee}&=&-\frac{i(\beta_1m_2-\beta_2m_1)}{8m_1m_2(m_1^2-m_2^2)c^6}
      [(\mathcal{OO}),(\mathcal{OE})(\mathcal{EO})] \qquad
   \label{Eq:9a}\\ \nonumber \\
      &&-\frac{i(\beta_1m_2+\beta_2m_1)}{8m_1m_2(m_1^2-m_2^2)c^6}
   \nonumber \\ \nonumber \\
      &&\times\Bigl\{
         (\mathcal{OE})(\mathcal{OO})(\mathcal{EO})-(\mathcal{EO})(\mathcal{OO})(\mathcal{OE})
      \Bigr\}.
   \label{Eq:9b}
\end{eqnarray}
\end{subequations}
Taking into account its higher order, one can retain only the first two terms in the series (\ref{Eq:2}), while the rest of the terms can be discarded because they are of lower order of magnitude,
\begin{equation}\label{Eq:10}
   e^{iS_{ee}}H_{\rm tr}e^{-iS_{ee}}\approx H_{\rm tr}+i[S_{ee},H_{\rm tr}].
\end{equation}
Since $S_{ee}$ has an even-even form, it commutes with the two large terms from $H_{\rm tr}\,$:
\begin{equation}\label{Eq:11}
   [S_{ee},\,\beta_1m_1c^2+\beta_2m_2c^2]=0.
\end{equation}
Actually, to remove the terms (\ref{Eq:8}), it is quite convenient to retain the members of order $c^0$ in $H_{\rm tr}$ standing in the commutator in (\ref{Eq:10}) and to omit the rest. It is sufficient to get the terms which coincide with the ones in Eq.~(\ref{Eq:8}) up to a sign. One sees that the operators (\ref{Eq:9a}) and (\ref{Eq:9b}) act separately. The former destroys the first, third, and fourth terms in Eq.~(\ref{Eq:8}), and the latter destroys the rest. We note that in general while removing the extra terms, this procedure gives rise to new ones instead, but all of them are of lower order of magnitude.

Thus the generating function $S_{ee}$ allows us to modify the higher-order transformed Hamiltonian, subtracting (or removing) the terms (\ref{Eq:8}). One can express it in terms of the original operators $S_{oe}$, $S_{eo}$, and $S_{oo}$ in a convenient brief form:
\begin{equation}\label{Eq:12}
   S_{ee}=[S_{oe},[S_{eo},S_{oo}]].
\end{equation}

In conclusion of this section we should point out that if the commutation relation (\ref{Eq:5}) had not been taken into account, the expression for $H_{\rm tr}$ would be much lengthier than that in Eq.~(\ref{Eq:6}) and many other terms would appear in it as well. Moreover, the form of $S_{ee}$ would be more complicated than we have in Eq.~(\ref{Eq:12}).

\section{\label{sec:3}Examples}
Our scope here is to apply the results obtained above to expand the Breit equation and the Salpeter equation to order $1/c^4$. For illustrative purposes, we consider the case of Coulomb particles with unequal masses and of charges $\epsilon_1$ and $\epsilon_2$, interacting through the potential
\begin{equation}\label{Eq:13}
   V({\bf r})=\frac{\epsilon_1\epsilon_2}{r}-\frac{\epsilon_1\epsilon_2}{2r}\left(
                   \boldsymbol{\alpha}_1\cdot\boldsymbol{\alpha}_2
                   +\frac{(\boldsymbol{\alpha}_1\cdot{\bf r})
                   (\boldsymbol{\alpha}_2\cdot{\bf r})}{r^2}
                \right),
\end{equation}
where ${\bf r}={\bf r}_1-{\bf r}_2$ and $r=|{\bf r}|$. One should remember, however, that in general the complete form of the original interaction of two Coulomb particles also has to include many other components of lower order of magnitude in addition to the Breit interaction (\ref{Eq:13}), e.g., such as the intrinsic magnetic moment terms \cite{barker}, the terms involving the electron self-energy and vacuum polarization \cite{indelicato}, etc. Nevertheless, here for simplicity we restrict ourselves to the consideration of an expansion of the equations only with the potential in the form (\ref{Eq:13}).

\subsection{\label{subsec:3.1}The Breit Equation}
Let us consider the Breit equation $H\psi=E\psi$, which Hamiltonian is similar to the one in Eq. (\ref{Eq:1}) if we put
\begin{subequations}\label{Eq:14}
\begin{eqnarray}
   &&(\mathcal{EE})=\cfrac{\epsilon_1\epsilon_2}{r}\,,         \label{Eq:14a}
      \\ \nonumber \\
   &&(\mathcal{OE})=c\,\boldsymbol{\alpha}_1\cdot{\bf p}_1,~~(\mathcal{EO})=c\,\boldsymbol{\alpha}_2\cdot{\bf p}_2, \label{Eq:14b}
      \\ \nonumber \\
   &&(\mathcal{OO})=-\cfrac{\epsilon_1\epsilon_2}{2r}
      \left(
          \boldsymbol{\alpha}_1\cdot\boldsymbol{\alpha}_2
          +\cfrac{(\boldsymbol{\alpha}_1\cdot{\bf r})(\boldsymbol{\alpha}_2\cdot{\bf r})}{r^2}
       \right).    \label{Eq:14c} \qquad \quad
\end{eqnarray}\end{subequations}
The even-even and odd-odd terms in this case denote the original interaction in $H$ and commute with each other. Using the expression (\ref{Eq:6a}), one performs the nonrelativistic expansion of the Breit equation to order $1/c^2$ and, putting $\beta_1=\beta_2=1$, gets the Breit correction derived also in QED (see, e.g., \cite{achiezerberestetskii,berestetskii}). Based on Eq.~(\ref{Eq:4}), which goes into (6a) in the case under consideration, an expansion of the Breit equation to order $1/c^2$ and the study of properties of its transformed Hamiltonian were carried out in~\cite{chraplyvy,barker}. Note that the Breit correction is divergent as it involves the Dirac $\delta$-functions appearing because of the Coulomb potential in the original Hamiltonian and because of the same potential, the $\delta$-functions, already together with their derivatives, will also appear and in the expansion terms in higher orders.

Since the prescription (\ref{Eq:6}) is just suitable for the case of the Breit equation, it is sufficient to work out the terms (\ref{Eq:6b}...j) with regard to the notations (\ref{Eq:14}), and thereby to get the $1/c^4$-order part of the transformed Hamiltonian in an explicit form for this equation. In order not to miss any $\delta$-functions and their derivatives in the final results, first one has to calculate these terms in momentum space and then to pass into coordinate space, as they usually do for the derivation of the Breit correction. Our main goal in this subsection is to obtain all the terms of order $1/c^4$ in the expansion of the Breit equation, which final forms we write below mainly in coordinate space.

We start off with the terms (\ref{Eq:6j})
\begin{equation}\label{Eq:15}
   \frac{\beta_1}{16m_1^5c^{10}}(\mathcal{OE})^6+\frac{\beta_2}{16m_2^5c^{10}}(\mathcal{EO})^6
         =\sum_{a=1,2}\frac{\beta_a {\bf p}_a^6}{16m_a^5c^{4}}\,,
\end{equation}
which provided that both particles are in positive energy states yield the correction of order $1/c^4$ to the kinetic energy. The rest of the members in (\ref{Eq:6}), except (\ref{Eq:6a}), lead to $1/c^4$-order corrections to the effective potential, form the interaction terms in the higher-order Breit-Fermi Hamiltonian, and with respect to the contribution of $(\mathcal{E}\mathcal{E})$ and $(\mathcal{O}\mathcal{O})$ can be divided into three groups.

The terms from the first group give corrections to the Coulomb interaction. They are represented by (\ref{Eq:6b},~g,~h). For the first term in (\ref{Eq:6b}), after substituting the operators from (\ref{Eq:14a},~b) and working out, one gets ($\hbar=1$)
\begin{equation}\label{Eq:16}
   [(\mathcal{O}\mathcal{E}), (\mathcal{E}\mathcal{E})]^2
      =-\,\frac{(\epsilon_1\epsilon_2)^2 c^2}{r^4}\,.
\end{equation}
The calculation for the sum of the two terms in the first bracket of (\ref{Eq:6g}) gives the following formula:
\begin{eqnarray}\label{Eq:17}
   &&[(\mathcal{OE}),[(\mathcal{OE}),[(\mathcal{OE}), [(\mathcal{OE}),(\mathcal{EE})]]]]
   \nonumber \\ \nonumber \\
   &&+32[(\mathcal{OE})^3,[(\mathcal{OE}),(\mathcal{EE})]]
   \nonumber \\ \nonumber \\
         &&~~~~~=72\epsilon_1\epsilon_2c^4\biggl\{
                    \pi[{\bf p}_1^2,\delta({\bf r})]_+
                    -2\pi(\nabla\delta({\bf r})\times{\bf p}_1)\cdot\boldsymbol{\sigma}_1
   \nonumber \\ \nonumber \\
         &&~~~~~~~~+\frac{({\bf r}\times{\bf p}_1)\cdot\boldsymbol{\sigma}_1}{r^3}{\bf p}_1^2
                    +3ir^i\frac{({\bf r}\times{\bf p}_1)\cdot\boldsymbol{\sigma}_1}{r^5}p_1^i
                 \biggr\}
   \nonumber \\ \nonumber \\
         &&~~~~~~~~+15c^4\biggl[
                       {\bf p}_1^2,\biggl[
                          {\bf p}_1^2,\frac{\epsilon_1\epsilon_2}{r}
                       \biggr]
                   \biggr].
\end{eqnarray}
\begin{widetext}
\noindent We can work out the commutator as
\begin{equation}\label{Eq:18}
   \left[
      {\bf p}_1^2,\left[
         {\bf p}_1^2,\frac{1}{r}
      \right]
   \right]
   =-4\pi\Delta\delta({\bf r})-16\pi i\nabla\delta({\bf r})\cdot{\bf p}_1
      +4\left[
         \frac{4\pi}{3}\delta({\bf r})\delta^{ij}+\frac{1}{r^3}\left(
            \delta^{ij}-3\frac{r^ir^j}{r^2}
         \right)
      \right]p_1^ip_1^j.
\end{equation}
The next two members in the second bracket in (\ref{Eq:6g}) are symmetrical to the ones derived above and can be obtained from them by interchanging the indices $1$ and $2$.

Then we have from the term (\ref{Eq:6h})
\begin{eqnarray}\label{Eq:19}
   [(\mathcal{OE}),[(\mathcal{OE}), [(\mathcal{EO}),[(\mathcal{EO}),(\mathcal{EE})]]]]
      &=&\int e^{i{\bf q}\cdot{\bf r}}\,\frac{4\pi\epsilon_1\epsilon_2c^4}{{\bf q}^2}
         \left(
            {\bf q}^2-2i\boldsymbol{\sigma}_1\cdot({\bf q}\times{\bf p}_1)
         \right)\left(
            {\bf q}^2+2i\boldsymbol{\sigma}_2\cdot({\bf q}\times{\bf p}_2)
         \right)\frac{d^{3}q}{(2\pi)^3}
   \nonumber \\ \nonumber \\
      &=&\epsilon_1\epsilon_2c^4
         \biggl\{
            -4\pi\Delta\delta({\bf r})
            -8\pi(\nabla\delta({\bf r})\times{\bf p}_1)\cdot\boldsymbol{\sigma}_1
            +8\pi(\nabla\delta({\bf r})\times{\bf p}_2)\cdot\boldsymbol{\sigma}_2
   \nonumber \\ \nonumber \\
            &&+4\left[
               \frac{4\pi}{3}\delta({\bf r})\delta^{ij}
               +\frac{1}{r^3}\!\left( \delta^{ij}-3\frac{r^ir^j}{r^2} \right)
            \right]
            (\boldsymbol{\sigma}_1\times{\bf p}_1)^i(\boldsymbol{\sigma}_2\times{\bf p}_2)^j
         \biggr\}.
\end{eqnarray}\end{widetext}
One can easily see that singular operators containing the Dirac $\delta$-function and its derivatives enter into Eqs.~(\ref{Eq:17}) and (\ref{Eq:19}). As already remarked, that is because the $(\mathcal{EE})$ term, in the original Hamiltonian, is represented by the Coulomb potential in the case considered here. Thus the terms (\ref{Eq:6b},~g,~h) with allowance for Eqs. (\ref{Eq:16}), (\ref{Eq:17}), (\ref{Eq:19}) (and for their symmetric equations) give relativistic corrections to the Coulomb interaction in the transformed Hamiltonian. With neglect of the odd-odd operator (\ref{Eq:14c}) in the Breit equation, only these terms form the higher-order transformed Hamiltonian, which can be applied to the equal-mass case as well. It should be noted that the term similar to (\ref{Eq:6h}) with (\ref{Eq:19}) was obtained for the effective $\alpha^6mc^2$ Hamiltonian and discussed in \cite{pachucki}.

The second group of the terms, which is a little larger than the preceding one, comes from (\ref{Eq:6c},~d,~i) and thus is responsible for corrections that are conditioned by the Breit operator, namely, by $(\mathcal{O}\mathcal{O})$.

The calculation of the second term from (\ref{Eq:6c}) gives us such a result:
\begin{widetext}
\begin{eqnarray}\label{Eq:20}
   [(\mathcal{OE}),(\mathcal{OO})]_+^2
      &=&\frac{(\epsilon_1\epsilon_2)^2c^2}{r}\left\{
         \frac{2}{r^3}
            -\frac{2}{r^3}({\bf r\times p}_1)\cdot\boldsymbol{\sigma}_1
            -\frac{3}{r^3}({\bf r\times p}_1)\cdot\boldsymbol{\sigma}_2
            -\frac{18}{r^3}\boldsymbol{\sigma}_1\cdot\boldsymbol{\sigma}_2
            +\left(
               \frac{7}{r^3}-\frac{4\pi}{3}\delta({\bf r})
            \right)\right.
   \nonumber \\ \nonumber \\
            &&\left.\times\left(
               3\delta^{ij}-\frac{r^ir^j}{r^2}
            \right)\sigma_1^i\sigma_2^j
            +\frac{1}{r}\left(
                \delta^{ij}+3\frac{r^ir^j}{r^2}
             \right)p_1^ip_1^j
             +\frac{2i}{r^3}\,{\bf r\cdot p}_1
         \right\}.
\end{eqnarray}
The terms of the next type from (\ref{Eq:6d}), which are represented by the first and third ones, have a more awkward form and the calculations of them are rather laborious and lead to lengthy expressions. However, we can avoid their direct calculation and simplify the procedure instead, using the following relation (for the first term):
\begin{equation*}
   [(\mathcal{OO}),[(\mathcal{OE}), [(\mathcal{OE}),(\mathcal{OO})]_+]_+]_+
       =2[(\mathcal{OE}),(\mathcal{OO})]_+^2+[(\mathcal{OE}),[(\mathcal{OE}),(\mathcal{OO})^2]].
\end{equation*}
Here the first member standing on the right-hand side is already obtained in (\ref{Eq:20}), and at the same time the second member is much simpler than the one on the left-hand side. The calculation of it yields
\begin{eqnarray}\label{Eq:21}
   [(\mathcal{OE}),[(\mathcal{OE}),(\mathcal{OO})^2]]
     &=&\frac{(\epsilon_1\epsilon_2)^2 c^2}{r}
      \left\{
         -\frac{3}{r^3}+\frac{32\pi}{3}\delta({\bf r})
         +\frac{6}{r^3}({\bf r\times p}_1)\cdot\boldsymbol{\sigma}_1
         +\frac{3}{r^3}({\bf r\times p}_1)\cdot\boldsymbol{\sigma}_2
         +\frac{7}{r^3}\boldsymbol{\sigma}_1\cdot\boldsymbol{\sigma}_2 \right.
   \nonumber \\ \nonumber \\
         &&-2\left(
                \frac{1}{r^3}+\frac{4\pi}{3}\delta({\bf r})
             \right)\left(
                3\delta^{ij}-\frac{r^ir^j}{r^2}
             \right)\sigma_1^i\sigma_2^j
         -\frac{2}{r}\left(2\delta^{ij}-\frac{r^ir^j}{r^2}\right)
            \left(
               \sigma_1^i\sigma_2^j{\bf p}_1^2
               -(\boldsymbol{\sigma}_1\cdot{\bf p}_1)\sigma_2^ip_1^j
            \right)
   \nonumber \\ \nonumber \\
         &&\left.+\frac{i}{r^3}\left[
               2r^ip_1^j+3r^jp_1^i-\left(7\delta^{ij}-4\frac{r^ir^j}{r^2}\right){\bf r\cdot p}_1
            \right]\sigma_1^i\sigma_2^j
      \right\}.
\end{eqnarray}
For the second type of the anticommutators from (\ref{Eq:6d}) we have
\begin{eqnarray}\label{Eq:22}
   [(\mathcal{OO}),[(\mathcal{OO}),(\mathcal{OE})^2]_+]_+
      &=&\frac{(\epsilon_1\epsilon_2)^2c^2}{r}\left\{
         -\frac{1}{r^3}+\frac{64\pi}{3}\delta({\bf r})
         +\frac{1}{r^3}({\bf r\times p}_1)\cdot
             \left(\boldsymbol{\sigma}_1+\boldsymbol{\sigma}_2\right)
         +\frac{13}{r^3}\boldsymbol{\sigma}_1\cdot\boldsymbol{\sigma}_2
         -4\left(
            \frac{1}{r^3}+\frac{4\pi}{3}\delta({\bf r})
         \right)\right.
   \nonumber \\ \nonumber \\
   &&\times\left(
            3\delta^{ij}-\frac{r^ir^j}{r^2}
         \right)\sigma_1^i\sigma_2^j
         +\frac{2}{r}\left[
            3-\left(
               2\delta^{ij}-\frac{r^ir^j}{r^2}
            \right)\sigma_1^i\sigma_2^j
         \right]{\bf p}_1^2+\frac{12i}{r^3}{\bf r\cdot p}_1
   \nonumber \\ \nonumber \\
   &&-\left.\frac{2i}{r^3}
         \left[
            r^ip_1^j+r^jp_1^i
            +4\left(
               \delta^{ij}-\frac{r^ir^j}{r^2}
            \right){\bf r\cdot p}_1
         \right]\sigma_1^i\sigma_2^j
      \right\}.
\end{eqnarray}
\end{widetext}

The two members in the first bracket in (\ref{Eq:6i}), which are linear in
$(\mathcal{O}\mathcal{O})$, $e^2$ terms, can be worked as follows:
\begin{eqnarray}\label{Eq:23}
   &&[(\mathcal{OE}),[(\mathcal{OE}), [(\mathcal{OE}),[(\mathcal{EO}),(\mathcal{OO})]_+]_+]_+]_+
   \nonumber \\ \nonumber \\
   &&+8[(\mathcal{OE})^3,[(\mathcal{EO}),(\mathcal{OO})]_+]_+
   \nonumber \\ \nonumber
   &&~~~=-\epsilon_1\epsilon_2 c^4\left(
      2p_2^j-(\boldsymbol{\sigma}_2\times\nabla)^j
   \right)\biggl\{
      \Bigl(
         2p_1^i+(\boldsymbol{\sigma}_1\times\nabla)^i
      \Bigr)
   \nonumber \\ \nonumber \\
   &&~~~~~~~\times\left[
         {\bf p}_1^2, \frac{3}{r}
         \left(
            \delta^{ij}+\frac{r^ir^j}{r^2}
         \right)
      \right]_+
      +(\boldsymbol{\sigma}_1\times(2i{\bf p}_1-\nabla))^i
   \nonumber \\ \nonumber \\
      &&~~~~~~~\left.\times\left[
         {\bf p}_1^2, \frac{3}{2r}
         \left(
            \delta^{ij}+\frac{r^ir^j}{r^2}
         \right)
      \right]
   \right\}.
\end{eqnarray}
The rest of the members from (\ref{Eq:6c},~d,~i) are symmetrical to the calculated ones.

Since the terms (\ref{Eq:6b}, g, h) involve the operators (\ref{Eq:14a},~b) only, their calculation gives relatively simple equations, however, it is well seen that the terms calculated for the second group have lengthier expressions than the terms with Eqs.~(\ref{Eq:16}), (\ref{Eq:17}), and (\ref{Eq:19}). The expressions obtained in Eqs.~(\ref{Eq:20})~--~(\ref{Eq:23}) have very cumbersome forms because the original terms (\ref{Eq:6c},~d,~i) contain six $\sigma$-matrices in our case and this leads to tedious calculations.

At last, the third group of the expansion terms comes from (\ref{Eq:6e}). All of them involve both the even-even and odd-odd operators and therefore give contributions that are conditioned by both the static Coulomb interaction and the Breit operator. They are nonlinear in the original interaction, thus, are $e^4$ terms.

The calculation of the first member in (\ref{Eq:6e}) yields
\begin{eqnarray}\label{Eq:24}
   &&[(\mathcal{OO}),[(\mathcal{OE}),[(\mathcal{EO}), (\mathcal{EE})]]]_+
   \nonumber \\ \nonumber \\
   &&~~~=\frac{2c^2}{(2\pi)^6}\iint d^3q d^3k\,\frac{4\pi\epsilon_1\epsilon_2}{{\bf q}^2}
         \frac{4\pi\epsilon_1\epsilon_2}{{\bf k}^2}
         \left(
            \delta^{ij}-\frac{k^ik^j}{{\bf k}^2}
         \right)
   \nonumber \\ \nonumber \\
   &&~~~~~~~\times\Bigl\{
                q^iq^j-(\boldsymbol{\sigma}_1\times{\bf q})^i
                (\boldsymbol{\sigma}_2\times{\bf q})^j
            \Bigr\}e^{i({\bf q}+{\bf k})\cdot{\bf r}}
   \nonumber \\ \nonumber \\
   &&~~~=\frac{2(\epsilon_1\epsilon_2)^2c^2}{r}
             \left\{
                -\frac{1}{r^3}+\frac{8\pi}{3}\delta({\bf r})
                -\frac{3}{r^3}\boldsymbol{\sigma}_1\cdot\boldsymbol{\sigma}_2 \right.
   \nonumber \\ \nonumber \\
   &&~~~~~~~+\left.\left(
                    \frac{1}{r^3}-\frac{2\pi}{3}\delta({\bf r})
                \right)\left(
                    3\delta^{ij}-\frac{r^ir^j}{r^2}
                \right)\sigma_1^i\sigma_2^j
             \right\}.~~~~~~~
\end{eqnarray}
The expression for the second member has the form
\begin{eqnarray}\label{Eq:25}
   &&[[(\mathcal{EO}),(\mathcal{EE})], [(\mathcal{OE}),(\mathcal{OO})]_+]
   \nonumber \\ \nonumber \\
   &&~~~~~=\frac{2(\epsilon_1\epsilon_2)^2c^2}{r}\left\{
       \frac{1}{r^3}-\frac{8\pi}{3}\delta({\bf r})
       -\frac{1}{r^3}({\bf r}\times{\bf p}_1)\cdot\boldsymbol{\sigma}_2\right.
   \nonumber \\ \nonumber \\
   &&~~~~~~~~\left.+\frac{1}{r^3}\left(
          \delta^{ij}-\frac{r^ir^j}{r^2}
       \right)\sigma_1^i\sigma_2^j
      \right\}.
\end{eqnarray}
Notice that all the terms from this group are only spin-depended ones. So far as the $e^6$ term (\ref{Eq:6f}) is concerned, it equals to zero because $(\mathcal{E}\mathcal{E})$ and $(\mathcal{O}\mathcal{O})$ commute in the case under consideration.

Thus, the terms (\ref{Eq:6b}...j) with regard to Eqs. (\ref{Eq:15})~--~(\ref{Eq:25}) form the $1/c^4$-order part of the transformed Hamiltonian for the Breit equation. However, as shown in the previous section, one can change it a little bit, applying a unitary transformation with the generator in the form (\ref{Eq:12}). Still, this transformation is not the only one that modifies the higher-order Hamiltonian.

Consider, for example, a Hermitian even-even operator in the following form \footnote{We use a common notation $S_{ee}$ for all the even-even generating functions considered.}:
\begin{equation}\label{Eq:26}
   S_{ee}=-\frac{i}{16c^6}\left[
      \frac{\beta_1(\mathcal{OE})^2}{m_1^3}+\frac{\beta_2(\mathcal{EO})^2}{m_2^3}, (\mathcal{EE})
   \right].
\end{equation}
We bear in mind here that the operators $(\mathcal{EE})$, $(\mathcal{OE})$, and $(\mathcal{EO})$ are defined by (\ref{Eq:14a}, b). Under this assumption the function (\ref{Eq:26}) reads
\begin{equation}\label{Eq:27}
   S_{ee}=-\frac{i}{16c^4}\left[
      \frac{\beta_1{\bf p}_1^2}{m_1^3}+\frac{\beta_2{\bf p}_2^2}{m_2^3},
          \frac{\epsilon_1\epsilon_2}{r}
   \right].
\end{equation}
Applying the procedure (\ref{Eq:2}) with the generating function in the form (\ref{Eq:27}) to the transformed Hamiltonian for the Breit equation, and with neglect of the terms of sixth and higher orders, one obtains
\begin{eqnarray}\label{Eq:28}
   (H_{\rm tr})'&\approx&H_{\rm tr}
      -\left(
         \frac{\beta_1}{m_1^3}+\frac{\beta_2}{m_2^3}
       \right)\frac{(\epsilon_1\epsilon_2)^2}{8c^4r^4}
      -\frac{1}{32m_1^4c^4}
   \nonumber \\ \nonumber \\
      &&\times\biggl[
         {\bf p}_1^2, \biggl[
            {\bf p}_1^2, \frac{\epsilon_1\epsilon_2}{r}
         \biggr]
       \biggr]
       -\frac{1}{32m_2^4c^4}\biggl[
         {\bf p}_2^2, \biggl[
            {\bf p}_2^2, \frac{\epsilon_1\epsilon_2}{r}
         \biggr]
       \biggr]
   \nonumber \\ \nonumber \\
      &&-\frac{\beta_1\beta_2(m_1^2+m_2^2)}{32m_1^3m_2^3c^4}
      \biggl[
          {\bf p}_1^2, \biggl[
             {\bf p}_2^2, \frac{\epsilon_1\epsilon_2}{r}
          \biggr]
      \biggr].
\end{eqnarray}
One can see that the term coming after $H_{\rm tr}$ is similar, up to a sign, to the term obtained from (\ref{Eq:6b}) (see Eq.~(\ref{Eq:16})), and therefore this additional transformation removes the last one from $H_{\rm tr}$ in the treated case. At the same time, given that the function (\ref{Eq:26}) is taken with another factor such as $-i5/(64c^6)$ instead of $-i/(16c^6)$, one destroys the last member in (\ref{Eq:17}), but the terms (\ref{Eq:6b}) will be saved out, however, with a different numerical multiplier.

Let us also consider two more generators:
\begin{subequations}\label{Eq:29}
\begin{eqnarray}
   S_{ee}=-\frac{i\beta_1}{16m_1m_2^2c^6}
      [[(\mathcal{O}\mathcal{E}),(\mathcal{O}\mathcal{O})]_+, (\mathcal{E}\mathcal{O})],&&   \label{Eq:29a}
   \\ \nonumber \\
   S_{ee}=-\frac{i\beta_2}{16m_1^2m_2c^6}
      [[(\mathcal{E}\mathcal{O}),(\mathcal{O}\mathcal{O})]_+, (\mathcal{O}\mathcal{E})].&&   \label{Eq:29b}
\end{eqnarray}
\end{subequations}
Here the second operator is symmetrical to the first one. They can also be applied to the transformed Hamiltonian to modify its higher-order part. Using the function in the form of (\ref{Eq:29b}), one obtains such new terms of order $1/c^4$ in the Hamiltonian:
\begin{subequations}\label{Eq:30}
\begin{eqnarray}
   i[S_{ee},(\mathcal{E}\mathcal{E})]
      &=&-\frac{\beta_2(\epsilon_1\epsilon_2)^2}{8m_1^2m_2c^4r^4}\biggl\{
         \left(\delta^{ij}-\frac{r^ir^j}{r^2}\right)\sigma_1^i\sigma_2^j
   \nonumber \\ \nonumber \\
       &&+2({\bf r\times p}_1)\cdot\boldsymbol{\sigma}_1
            +({\bf r\times p}_2)\cdot\boldsymbol{\sigma}_1
      \biggr\},~~~~~~                                              \label{Eq:30a}
\end{eqnarray}
\begin{eqnarray}
   i\left[S_{ee},\frac{\beta_1(\mathcal{O}\mathcal{E})^2}{2m_1c^2}\right]
      &=&\frac{\beta_1\beta_2\epsilon_1\epsilon_2}{32m_1^3m_2c^4}
         (\boldsymbol{\sigma}_1\times(2i{\bf p}_1-\nabla))^i
   \nonumber \\ \nonumber \\
      &&\times\left(
            2p_2^j-(\boldsymbol{\sigma}_2\times\nabla)^j
        \right)
   \nonumber \\ \nonumber \\
      &&\times\left[
      {\bf p}_1^2, -\frac{1}{2r}\left(\delta^{ij}+\frac{r^ir^j}{r^2}\right)
   \right],~~~~~~~~                                                           \label{Eq:30b}
\end{eqnarray}
\begin{eqnarray}
   i\left[S_{ee},\frac{\beta_2(\mathcal{E}\mathcal{O})^2}{2m_2c^2}\right]
      &=&\frac{\epsilon_1\epsilon_2}{32m_1^2m_2^2c^4}
         (\boldsymbol{\sigma}_1\times(2i{\bf p}_1-\nabla))^i
   \nonumber \\ \nonumber \\
      &&\times\left(
            2p_2^j-(\boldsymbol{\sigma}_2\times\nabla)^j
        \right)
   \nonumber \\ \nonumber \\
      &&\times\left[
      {\bf p}_2^2, -\frac{1}{2r}\left(\delta^{ij}+\frac{r^ir^j}{r^2}\right)
   \right].~~~~~~~~                                                          \label{Eq:30c}
\end{eqnarray}
\end{subequations}
We have in (\ref{Eq:30b}) the expression that is opposite to the sign of the second term in the bracket in (\ref{Eq:23}). The transformation with the generating function (\ref{Eq:29a}) leads to a similar modification of the transformed Hamiltonian but in its symmetrical part.

Thus the higher-order Breit-Fermi Hamiltonian can be modified by the unitary transformations. We considered here these transformations with even-even functions that contain the operators determined by (\ref{Eq:14}), but in general they may be other functions of ${\bf r}$, ${\bf p}_1$, ${\bf p}_2$, $\boldsymbol{\sigma}_1$, $\boldsymbol{\sigma}_2$ as well. With due regard for transformations of this kind, one can destroy (or create) a number of higher-order terms in the transformed Hamiltonian. Meanwhile, as shown in the examples considered these procedures produce, as a rule, new terms instead of the removed ones. Note that these transformations only change the higher-order part of the approximate Hamiltonian.

Obviously, using the same procedure with ``even'' generators, one can easily modify the transformed one-body Dirac Hamiltonian (\ref{Eq:7}) in a similar way.

\subsection{\label{subsec:3.2}The Salpeter Equation}
Consider the Hamiltonian of the Hermitian part of the three-dimensional Bethe-Salpeter equation in coordinate space \cite{salpeter}:
\begin{equation}\label{Eq:31}
      H=H_1+H_2+\frac{1}{2}\left[
         \left(
            \Lambda_1^+\Lambda_2^+-\Lambda_1^-\Lambda_2^-
         \right),
         V({\bf r})
      \right]_+,
\end{equation}
where
\begin{eqnarray*}
   &H_a=c\,\boldsymbol{\alpha}_a\cdot{\bf p}_a+\beta_am_ac^2,\quad
     \Lambda_a^{\pm}=\cfrac{E_a\pm H_a}{2E_a}\;,& \\ \\
   &E_a=\left(m_a^2c^4+{\bf p}_a^2c^2\right)^{1/2},&
\end{eqnarray*}
with $a=1,2$, and $V({\bf r})$ is determined by (\ref{Eq:13}). In order to reduce this Hamiltonian to the form (\ref{Eq:1}), we evaluate the anticommutator, keeping the terms which contribute to order $1/c^4$ in the expansion. Hence we have
\begin{eqnarray*}
   (\mathcal{E}\mathcal{E})&=&(\beta_1+\beta_2)\frac{\epsilon_1\epsilon_2}{2r}
      -\left[
          \frac{\beta_1{\bf p}_1^2}{8m_1^2c^2}+\frac{\beta_2{\bf p}_2^2}{8m_2^2c^2},
                \frac{\epsilon_1\epsilon_2}{r}
       \right]_{+}
   \\ \\
      &&+\left[
            \frac{3\beta_1{\bf p}_1^4}{32m_1^4c^4}+\frac{3\beta_2{\bf p}_2^4}{32m_2^4c^4},
               \frac{\epsilon_1\epsilon_2}{r}
         \right]_{+},
   \\ \\ \\
   (\mathcal{O}\mathcal{E})&=&c\,\boldsymbol{\alpha}_1\cdot{\bf p}_1
      +\left[
          \frac{\boldsymbol{\alpha}_1\cdot{\bf p}_1}{4m_1c}
             -\frac{(\boldsymbol{\alpha}_1\cdot{\bf p}_1){\bf p}_1^2}{8m_1^3c^3},
                \frac{\epsilon_1\epsilon_2}{r}
       \right]_{+}
   \\   \\
      &&-\left[
            \frac{\boldsymbol{\alpha}_2\cdot{\bf p}_2}{4m_2c},
                \frac{\epsilon_1\epsilon_2}{2r}\left(
                   \delta^{ij}+\frac{r^ir^j}{r^2}
               \right)\alpha_1^i\alpha_2^j
         \right]_{+}
    \\  \\
      &&+\left[
           \frac{(\boldsymbol{\alpha}_2\cdot{\bf p}_2){\bf p}_2^2}{8m_2^3c^3},
              \frac{\epsilon_1\epsilon_2}{2r}\left(
                 \delta^{ij}+\frac{r^ir^j}{r^2}
              \right)\alpha_1^i\alpha_2^j
         \right]_{+},
    \\ \\  \\
  (\mathcal{E}\mathcal{O})&=&c\,\boldsymbol{\alpha}_2\cdot{\bf p}_2
      +\left[
          \frac{\boldsymbol{\alpha}_2\cdot{\bf p}_2}{4m_2c}
             -\frac{(\boldsymbol{\alpha}_2\cdot{\bf p}_2){\bf p}_2^2}{8m_2^3c^3},
                 \frac{\epsilon_1\epsilon_2}{r}
       \right]_{+}
   \nonumber \\ \nonumber \\
      &&-\left[
            \frac{\boldsymbol{\alpha}_1\cdot{\bf p}_1}{4m_1c},
                \frac{\epsilon_1\epsilon_2}{2r}\left(
                    \delta^{ij}+\frac{r^ir^j}{r^2}
                \right)\alpha_1^i\alpha_2^j
         \right]_{+}
   \\ \\
      &&+\left[
             \frac{(\boldsymbol{\alpha}_1\cdot{\bf p}_1){\bf p}_1^2}{8m_1^3c^3},
                \frac{\epsilon_1\epsilon_2}{2r}\left(
                    \delta^{ij}+\frac{r^ir^j}{r^2}
                \right)\alpha_1^i\alpha_2^j
         \right]_{+},
  \\  \\  \\
   (\mathcal{O}\mathcal{O})&=&
      \left[
          \frac{\beta_1{\bf p}_1^2}{8m_1^2c^2}+\frac{\beta_2{\bf p}_2^2}{8m_2^2c^2},
             \frac{\epsilon_1\epsilon_2}{2r}\left(
                 \delta^{ij}+\frac{r^ir^j}{r^2}
             \right)\alpha_1^i\alpha_2^j
       \right]_{+}.
\end{eqnarray*}

It is seen that to perform an expansion of the Salpeter equation, one has to use a prescription in a general form like (\ref{Eq:4}), as the operators $(\mathcal{O}\mathcal{E})$ and $(\mathcal{E}\mathcal{O})$ do not commute in the case under consideration. Still, if we supplement expression (\ref{Eq:6}) with the terms (\ref{Eq:4h}, i), we will get a prescription for the transformation of the Hamiltonian (\ref{Eq:31}) to fourth order in $1/c$. It should be noted, however, that as $(\mathcal{O}\mathcal{O})$ is of order $1/c^2$, the terms containing this operator, except the ninth term in (\ref{Eq:6a}), make no contribution to the desired accuracy, and hence, can be ignored (all of the terms with a mass difference in the denominators are among them). Thus, after the evaluation, one obtains the transformed Hamiltonian from the Salpeter equation to order $1/c^4$ in the following form:
\begin{widetext}
\begin{subequations} \label{Eq:32}
\begin{eqnarray}
   H_{\rm tr}&\approx&\beta_1m_1c^2+\frac{\beta_1{\bf p}_1^2}{2m_1}+(\beta_1+\beta_2)
      \frac{\epsilon_1\epsilon_2}{4r}
      -\frac{\beta_1+\beta_2}{16m_1^2c^2}
          \biggl[(\boldsymbol{\sigma}_1\cdot{\bf p}_1),
             \biggl[
                (\boldsymbol{\sigma}_1\cdot{\bf p}_1),\frac{\epsilon_1\epsilon_2}{r}
             \biggr]
          \biggr]      \nonumber
   \\ \nonumber \\
      &&-\frac{\beta_1+\beta_2}{16m_1m_2c^2}
         \biggl[
            (\boldsymbol{\sigma}_1\cdot{\bf p}_1),\left[(\boldsymbol{\sigma}_2\cdot{\bf p}_2),
               \frac{\epsilon_1\epsilon_2}{2r}
               \left(
                  \delta^{ij}+\frac{r^ir^j}{r^2}
               \right)
               \sigma_1^i\sigma_2^j\right]_{+}\biggr]_{+}
        -\frac{\beta_1{\bf p}_1^4}{8m_1^3c^2}               \label{Eq:32a}
   \\ \nonumber \\
      &&-\frac{\beta_1}{32m_1^3c^4}
           \biggl[
               (\boldsymbol{\sigma}_1\cdot{\bf p}_1),\frac{\epsilon_1\epsilon_2}{r}
           \biggr]^2
        +\frac{\beta_1}{32m_1m_2^2c^4}
           \left[
              (\boldsymbol{\sigma}_2\cdot{\bf p}_2),\frac{\epsilon_1\epsilon_2}{2r}
              \left(
                 \delta^{ij}+\frac{r^ir^j}{r^2}
              \right)
              \sigma_1^i\sigma_2^j\right]_{+}^2      \nonumber
   \\ \nonumber \\
      &&+\frac{\beta_1}{32m_1m_2^2c^4}
           \biggl[
              \left[
                 (\boldsymbol{\sigma}_1\cdot{\bf p}_1),\frac{\epsilon_1\epsilon_2}{2r}
                 \left(
                    \delta^{ij}+\frac{r^ir^j}{r^2}
                 \right)
                 \sigma_1^i\sigma_2^j
              \right]_{+},
              \biggl[
                 (\boldsymbol{\sigma}_2\cdot{\bf p}_2),\frac{\epsilon_1\epsilon_2}{r}
              \biggr]
          \biggr]                          \label{Eq:32b}
   \\ \nonumber \\
      &&+\frac{\beta_1+\beta_2}{768m_1^4c^4}
         \left\{
            \biggl[
               (\boldsymbol{\sigma}_1\cdot{\bf p}_1),
               \biggl[
                  (\boldsymbol{\sigma}_1\cdot{\bf p}_1),
                  \biggl[
                     (\boldsymbol{\sigma}_1\cdot{\bf p}_1),
                     \biggl[
                        (\boldsymbol{\sigma}_1\cdot{\bf p}_1),\frac{\epsilon_1\epsilon_2}{r}
                     \biggr]
                  \biggr]
               \biggr]
            \biggr]
            +32\biggl[
               (\boldsymbol{\sigma}_1\cdot{\bf p}_1),
                  \biggl[
                     (\boldsymbol{\sigma}_1\cdot{\bf p}_1){\bf p}_1^2,
                        \frac{\epsilon_1\epsilon_2}{r}
                  \biggr]
               \biggr]
         \right\}             \nonumber
   \\ \nonumber \\
      &&+\frac{\beta_1+\beta_2}{256m_1^2m_2^2c^4}
            \biggl[
               (\boldsymbol{\sigma}_1\cdot{\bf p}_1),
               \biggl[
                  (\boldsymbol{\sigma}_1\cdot{\bf p}_1),
                  \biggl[
                     (\boldsymbol{\sigma}_2\cdot{\bf p}_2),
                     \biggl[
                        (\boldsymbol{\sigma}_2\cdot{\bf p}_2),\frac{\epsilon_1\epsilon_2}{r}
                     \biggr]
                  \biggr]
               \biggr]
            \biggr]          \nonumber
   \\ \nonumber \\
      &&+\frac{\beta_1+\beta_2}{192m_1m_2^3c^4}
         \biggl\{
            \biggl[
               (\boldsymbol{\sigma}_1\cdot{\bf p}_1),
               \biggl[
                  (\boldsymbol{\sigma}_2\cdot{\bf p}_2),
                  \biggl[
                  (\boldsymbol{\sigma}_2\cdot{\bf p}_2),
                     \biggl[
                        (\boldsymbol{\sigma}_2\cdot{\bf p}_2),\frac{\epsilon_1\epsilon_2}{2r}
                        \left(
                           \delta^{ij}+\frac{r^ir^j}{r^2}
                        \right)
                        \sigma_1^i\sigma_2^j
                     \biggr]_+
                  \biggr]_+
               \biggr]_+
            \biggr]_+                  \nonumber
   \\ \nonumber \\
      &&~~~~~~~~~~~~~~~~~~~~~+8\biggl[
             (\boldsymbol{\sigma}_1\cdot{\bf p}_1),
             \biggl[
                (\boldsymbol{\sigma}_2\cdot{\bf p}_2){\bf p}_2^2,\frac{\epsilon_1\epsilon_2}{2r}
                \left(
                   \delta^{ij}+\frac{r^ir^j}{r^2}
                \right)
                \sigma_1^i\sigma_2^j
             \biggr]_+
          \biggr]_+
       \biggr\}                        \label{Eq:32c}
   \\ \nonumber \\
      &&-\frac{\beta_1+\beta_2}{64m_1^3c^4}
            \biggl[
               \biggl[
                  {\bf p}_1^2,\frac{\epsilon_1\epsilon_2}{r}
               \biggr],
               \frac{\epsilon_1\epsilon_2}{r}
            \biggr]
        +\frac{\beta_1+\beta_2}{64m_1m_2^2c^4}
            \biggl[
               \biggl[
                  (\boldsymbol{\sigma}_1\cdot{\bf p}_1),
                  \left[
                     (\boldsymbol{\sigma}_2\cdot{\bf p}_2),\frac{\epsilon_1\epsilon_2}{2r}
                     \left(
                        \delta^{ij}+\frac{r^ir^j}{r^2}
                     \right)
                     \sigma_1^i\sigma_2^j\right]
               \biggr]_{+},
               \frac{\epsilon_1\epsilon_2}{r}
            \biggr]         \nonumber
   \\ \nonumber \\
      &&-\frac{1}{32m_1^3c^4}
            \biggl[
               \biggl[
                  {\bf p}_1^2,\frac{\epsilon_1\epsilon_2}{r}
               \biggr],
               \frac{\beta_1{\bf p}_1^2}{2m_1}+\frac{\beta_2{\bf p}_2^2}{2m_2}
            \biggr]           \nonumber
   \\ \nonumber \\
      &&+\frac{1}{32m_1m_2^2c^4}
         \biggl[
            \biggl[
               (\boldsymbol{\sigma}_1\cdot{\bf p}_1),
               \biggl[
                  (\boldsymbol{\sigma}_2\cdot{\bf p}_2),\frac{\epsilon_1\epsilon_2}{2r}
                  \left(
                     \delta^{ij}+\frac{r^ir^j}{r^2}
                  \right)
                  \sigma_1^i\sigma_2^j
               \biggr]
            \biggr]_{+},
            \frac{\beta_1{\bf p}_1^2}{2m_1}+\frac{\beta_2{\bf p}_2^2}{2m_2}
         \biggr]                                                             \label{Eq:32d}
   \\ \nonumber \\
      &&+\frac{\beta_1{\bf p}_1^6}{16m_1^5c^4}+{\rm symm.~terms}.            \label{Eq:32e}
   \\ \nonumber
\end{eqnarray}
\end{subequations}
\end{widetext}
Although this formula is only an intermediate result, the Hamiltonian written in such a form is quite transparent and suitable for a qualitative analysis. Besides, it is represented in the form similar to the one in (\ref{Eq:6}) with regard to the notations (\ref{Eq:14}), therefore, is convenient to compare with the transformed Hamiltonian of the corresponding Breit equation.

Indeed, expressing Eq.~(\ref{Eq:32}) in terms of the $(\mathcal{E}\mathcal{E})$, $(\mathcal{O}\mathcal{E})$, $(\mathcal{E}\mathcal{O})$, and $(\mathcal{O}\mathcal{O})$ operators from (\ref{Eq:14}), we can note that it is almost similar to equation (\ref{Eq:6}) \footnote{Here, we mean by Eq. (\ref{Eq:6}) the transformed Hamiltonian for the Breit equation.}. The terms in (\ref{Eq:32}), except the terms (\ref{Eq:32d}), are similar to the ones entering into the transformed Hamiltonian of the Breit equation. The $e^4$ terms, which we put together in (\ref{Eq:32b}) and which are of order $1/m^3$, do not disappear when one particle is in a positive energy state and the other is in a negative energy state. The similar terms together with their symmetrical ones also occur among the nonlinear terms (\ref{Eq:6b},~c,~e). Note, all of them agree with those $e^4$ terms in the expansion of the Breit equation, which do not have a mass difference in the denominators, however, they are four times higher. Meanwhile, the $e^2$ terms (\ref{Eq:32c}), which are of order $1/m^4$, exhibit another behavior. They have the factor $\beta_1+\beta_2$, and thus, become zero if one particle is in a positive energy state and the other is in a negative energy state. When both particles are in positive energy states, they coincide with the terms (\ref{Eq:6g},~h,~i) for the case of the Breit equation, but all of them are opposite in sign to those provided that the two particles are in negative energy states. As concerns the terms (\ref{Eq:32d}), there are no similar ones in Eq.~(\ref{Eq:6}). One should remark that this part includes $e^4$ terms having the factor $\beta_1+\beta_2$ (the $e^4$ terms of (\ref{Eq:32b}) are without this factor) and $e^2$ terms which do not disappear when the particles are in different energy states, as opposed to the terms (\ref{Eq:32c}).

It is easy to check, however, that such terms can be removed by an additional unitary transformation like those considered in the previous subsection. Namely, applying the procedure (\ref{Eq:2}) with the generator
\begin{widetext}
\begin{eqnarray} \label{Eq:33}
   S&=&-\,\frac{i}{32c^4}
           \left[
              \frac{{\bf p}_1^2}{m_1^3}+\frac{{\bf p}_2^2}{m_2^3}, \frac{\epsilon_1\epsilon_2}{r}
           \right]
       +\frac{i}{32m_1m_2^2c^4}
            \biggl[
               (\boldsymbol{\sigma}_1\cdot{\bf p}_1),
               \biggl[
                  (\boldsymbol{\sigma}_2\cdot{\bf p}_2),\frac{\epsilon_1\epsilon_2}{2r}
                  \left(
                     \delta^{ij}+\frac{r^ir^j}{r^2}
                  \right)
                  \sigma_1^i\sigma_2^j
               \biggr]
            \biggr]_{+}   \nonumber
   \\ \nonumber \\
       &&+\,\frac{i}{32m_1^2m_2c^4}
            \biggl[
               (\boldsymbol{\sigma}_1\cdot{\bf p}_1),
               \biggl[
                  (\boldsymbol{\sigma}_2\cdot{\bf p}_2),\frac{\epsilon_1\epsilon_2}{2r}
                  \left(
                     \delta^{ij}+\frac{r^ir^j}{r^2}
                  \right)
                  \sigma_1^i\sigma_2^j
               \biggr]_{+}
            \biggr] \\ \nonumber \\ \nonumber
\end{eqnarray}
\end{widetext}
to Eq.~(\ref{Eq:32}), one eliminates all the terms of (\ref{Eq:32d}) together with their symmetrical terms. This transformation only destroys the terms (\ref{Eq:32d}), without changing the rest part of the approximate Hamiltonian and without producing any new terms to the desired accuracy. In fact, this generator contains three functions similar to (\ref{Eq:27}) and (\ref{Eq:29}), but with other factors. Here the first function cancels the first and third terms from (\ref{Eq:32d}) with their symmetrical ones, and the next two functions cancel the rest. So, with neglect of the terms (\ref{Eq:32d}), $H_{\rm tr}$, in the case under consideration, involves only the terms similar in their structure to those occurring in the transformed Hamiltonian for the Breit equation.

Using the results obtained above, we can thus represent the final form of the transformed Hamiltonian for the Salpeter equation, accurate to order $1/c^4$, as
\begin{widetext}
\begin{subequations} \label{Eq:34}
\begin{eqnarray}
   H_{\rm tr}&\approx&\beta_1m_1c^2+\frac{\beta_1{\bf p}_1^2}{2m_1}
      +(\beta_1+\beta_2)\frac{\epsilon_1\epsilon_2}{4r}
      -(\beta_1+\beta_2)\frac{\epsilon_1\epsilon_2}{8m_1^2 c^2} \left(
            2\pi\delta({\bf r})+\frac{1}{r^3}\left({\bf r}
                 \times{\bf p}_1\right)\cdot\boldsymbol{\sigma}_1
         \right)
      -(\beta_1+\beta_2)\frac{\epsilon_1\epsilon_2}{16m_1 m_2 c^2}   \nonumber
   \\ \nonumber \\
      &&\times\left\{
         \frac{2}{r}\left(
             \delta^{ij}+\frac{r^ir^j}{r^2}
         \right)p_1^ip_2^j
         +\frac{4}{r^3}\left({\bf r}\times{\bf p}_1\right)\cdot\boldsymbol{\sigma}_2
         -\frac{1}{r^3}\left(
             \delta^{ij}-3\frac{r^ir^j}{r^2}
         \right)\sigma_1^i\sigma_2^j
         +\frac{8\pi}{3}\delta({\bf r})\boldsymbol{\sigma}_1\cdot\boldsymbol{\sigma}_2
      \right\}
      -\frac{\beta_1{\bf p}_1^4}{8m_1^3c^2}     \label{Eq:34a}
   \\ \nonumber \\
      &&+\frac{\beta_1(\epsilon_1\epsilon_2)^2}{32m_1^3c^4r^4}
      +\frac{\beta_1(\epsilon_1\epsilon_2)^2}{16m_1m_2^2c^4r}\left\{
         \frac{2}{r^3}-\frac{8\pi}{3}\delta({\bf r})
         +\frac{1}{r^3}\left(
            \left({\bf r}\times{\bf p}_2\right)\cdot\boldsymbol{\sigma}_2
            +\frac{3}{2}\left({\bf r}\times{\bf p}_2\right)\cdot\boldsymbol{\sigma}_1
            -\left({\bf r}\times{\bf p}_1\right)\cdot\boldsymbol{\sigma}_2
         \right)\right.                                                        \nonumber
   \\ \nonumber \\
         &&\left.
         -\frac{11}{r^3}\boldsymbol{\sigma}_1\cdot\boldsymbol{\sigma}_2
         +\left(
             \frac{9}{2r^3}-\frac{2\pi}{3}\delta({\bf r}) 
         \right)\left(
             3\delta^{ij}-\frac{r^ir^j}{r^2}
         \right)\sigma_1^i\sigma_2^j
         +\frac{1}{2r}\left(
              \delta^{ij}+3\frac{r^ir^j}{r^2}
         \right)p_2^ip_2^j
         -\frac{i}{r^3}{\bf r}\cdot{\bf p}_2
      \right\}                                   \label{Eq:34b}
   \\ \nonumber \\
      &&+(\beta_1+\beta_2)\frac{\epsilon_1\epsilon_2}{256m_1^4c^4}\biggl\{
         6\left[
             {\bf p}_1^2, 4\pi\delta({\bf r})
                +\frac{2}{r^3}({\bf r}\times{\bf p}_1)\cdot\boldsymbol{\sigma}_1
         \right]_{+}
         +5\left[{\bf p}_1^2, \left[{\bf p}_1^2, \frac{1}{r}\right]\right] 
      \biggr\}
      +(\beta_1+\beta_2)\frac{\epsilon_1\epsilon_2}{256m_1^2m_2^2c^4}       \nonumber
   \\ \nonumber \\
      &&\times\biggl\{
         -4\pi\Delta\delta({\bf r})-8\pi(\nabla\delta({\bf r})\times{\bf p}_1)
             \cdot\boldsymbol{\sigma}_1
         +8\pi(\nabla\delta({\bf r})\times{\bf p}_2)\cdot\boldsymbol{\sigma}_2
         +4\biggl[
             \frac{4\pi}{3}\delta({\bf r})\delta^{ij}
             +\frac{1}{r^3}\left(
                \delta^{ij}-3\frac{r^ir^j}{r^2}
             \right)                                        \nonumber
   \\ \nonumber \\
             &&\times\left(\boldsymbol{\sigma}_1\cdot{\bf p}_1\right)^i
             \left(\boldsymbol{\sigma}_2\cdot{\bf p}_2\right)^j
         \biggr]
      \biggr\}
      +\frac{\beta_1+\beta_2}{64m_1m_2^3c^4}
      \left(2p_1^i+\left(\boldsymbol{\sigma}_1\times\nabla\right)^i\right)
      \biggl\{
         \left(2p_2^j-\left(\boldsymbol{\sigma}_2\times\nabla\right)^j\right)
                \left[
                   {\bf p}_2^2, \frac{\epsilon_1\epsilon_2}{r}\left(
                      \delta^{ij}+\frac{r^ir^j}{r^2}
                   \right)
                \right]_{+}                           \nonumber
   \\ \nonumber \\
         &&+\Bigl(
             \boldsymbol{\sigma}_2\times\left(2i{\bf p}_2+\nabla\right)
         \Bigr)^j\left[
             {\bf p}_2^2, \frac{\epsilon_1\epsilon_2}{2r}\left(
                 \delta^{ij}+\frac{r^ir^j}{r^2}
             \right)
         \right]
      \biggr\}                   \label{Eq:34c}
  \\ \nonumber \\
      &&+\frac{\beta_1 {\bf p}_1^6}{16m_1^5c^4}+{\rm symm.~terms}.      \label{Eq:34d}
\end{eqnarray}
\end{subequations}
Here the terms to second order are put together in (\ref{Eq:34a}), and the others form the $1/c^4$-order part of the transformed Hamiltonian. We do not include the terms coming from (\ref{Eq:32d}) in this equation and represent them apart:
\begin{eqnarray} \label{Eq:35}
      &&(\beta_1+\beta_2)\frac{(\epsilon_1\epsilon_2)^2}{32m_1^3c^4r^4}
      +(\beta_1+\beta_2)\frac{(\epsilon_1\epsilon_2)^2}{32m_1m_2^2c^4r^4}
         \left\{
            2\left({\bf r}\times{\bf p}_2\right)\cdot\boldsymbol{\sigma}_2
            +\left({\bf r}\times{\bf p}_1\right)\cdot\boldsymbol{\sigma}_2
            -\left(
               \delta^{ij}-\frac{r^ir^j}{r^2}
            \right)
            \sigma_1^i\sigma_2^j
         \right\}                           \nonumber
   \\ \nonumber \\
      &&+\frac{\epsilon_1\epsilon_2}{32m_1^3c^4}
            \left[
               \frac{\beta_1{\bf p}_1^2}{2m_1}+\frac{\beta_2{\bf p}_2^2}{2m_2},
                  \left[
                     {\bf p}_1^2, \frac{1}{r}
                  \right]
            \right]
        -\frac{\epsilon_1\epsilon_2}{64m_1m_2^2c^4}
              \left(
                 2p_1^i+(\boldsymbol{\sigma}_1\times\nabla)^i
              \right)
              \bigl(
                 \boldsymbol{\sigma}_2\times\left(2i{\bf p}_2+\nabla\right)
              \bigr)^j           \nonumber
   \\ \nonumber \\
              &&\times\left[
                  \frac{\beta_1{\bf p}_1^2}{2m_1}+\frac{\beta_2{\bf p}_2^2}{2m_2},
                      \frac{1}{r}\left(
                          \delta^{ij}+\frac{r^ir^j}{r^2}
                      \right)
              \right]+{\rm symm.~terms}.
\end{eqnarray}
\end{widetext}
The extension of the FW method applied to the Salpeter equation here gives the formula with these terms, which can be referred to extra terms. Yet, we cannot claim that they also arise in the nonrelativistic expansion obtained by a different method, e.g., by the method of expressing the small components of the spinor of the wave equation through its large components \cite{BetheSalpeter,chraplyvy3}.

\section{\label{sec:4}Summary}
In the present paper we have dealt with the extension of the FW method to a system of two spin $1/2$ particles and, by means of which, we have briefly considered the problem of expansion of equal-time relativistic two-body wave equations to higher orders in $1/c$. For the case of unequal masses, we have found the fourth-order part of the transformed Hamiltonian in a general form in terms of even-even, even-odd, odd-even, and odd-odd operators. It turned out that in contrast with the case of the transformation to second order, the form of the higher-order transformed Hamiltonian depends on the order of application of the generators, i.e., as these functions do not commute with each other \footnote{To be exact, $S_{oe}$ and $S_{eo}$ commute with each other in the case of the Breit equation but do not commute with $S_{oo}$.}, certain extra terms can arise in the expansion of the original equation. Still, terms of this kind can be removed by an additional unitary transformation, which, however, does not change the transformed Hamiltonian to order $1/c^2$. The occurrence of extra terms in $H_{\rm tr}$ may be a feature of the expansion of relativistic equations to higher orders. Evidently, any of the model effective Hamiltonians giving a contribution to $\alpha^6mc^2$ to the energy levels of an arbitrary light atom, not only the approximate Hamiltonians that one obtains from the nonrelativistic reduction of the wave equations, is defined up to a unitary transformation with generating functions of order $1/c^4$.

For illustration, we have considered the nonrelativistic expansion of the Breit equation and the Hermitian part of the three-dimensional Bethe-Salpeter equation with the Breit interaction up to the terms of fourth order. Due to the Casimir projection operators, the $1/c^4$-order part of the transformed Hamiltonian obtained from the Salpeter equation is much simpler than the analogous part from the  reduction of the corresponding Breit equation, and does not involve any terms with mass differences in the denominators or terms similar to those in (\ref{Eq:6d}) by their structure. However, none of the $e^4$ terms from the expansion of the Salpeter equation (\ref{Eq:34}) disappears when one particle is in a positive energy state and the other is in a negative energy state, meanwhile all the $e^2$ terms become zero \footnote{It should be pointed out that all the interaction terms to order $1/c^2$ in the transformed Hamiltonian derived from the Salpeter equation have the factor $\beta_1+\beta_2$, however, all of them are $e^2$ terms and vanish when one particle is in a positive energy state and the other is in a negative energy state (see also Ref. \cite{barker}).}. Notice that a number of the $1/c^4$-order expansion terms worked out for these equations agree up to a numerical factor, and in many features, are similar to the relativistic corrections in the $\alpha^6mc^2$ Hamiltonian for an arbitrary light atom (see, e.g., \cite{pachucki}).

In fact, the approach applied here to expand the wave equations is a straightforward way of deriving many $1/c^4$-order terms giving $\alpha^6mc^2$ corrections to the energy levels for a hydrogen-like atom. We emphasize, however, that this work has been devoted mainly to an extension of the FW canonical transformation to the two-body problem. Although we have illustrated the application of the results on examples of the Breit and Salpeter equations, the application to a real atomic system and calculation of the energy states are beyond the purpose of the paper. Because of the simple form of the original interaction we used in the equations, the obtained approximate Hamiltonians are not complete and do not take into account many QED effects. The derivation of the total effective Hamiltonian contributing to $\alpha^6mc^2$ to the energy for a hydrogen-like atom by means of the considered method also requires the use of quantum field approaches and the diagram technique, which is of interest in itself and could be the subject of future investigations.

The results of the work are applicable to any two-body system, besides, they can be generalized to the systems of three and more number of particles. They may be useful in the study of few-body systems when the application of quantum field theories causes difficulties and can also be of some interest from the theoretical and mathematical points of view in general.


\begin{acknowledgments}
I would like to express my thanks to my colleagues for interesting and useful discussions, and I am also grateful to Professor I.~V.~Simenog for his critical comments.
\end{acknowledgments}


\end{document}